# Nonmonotonic In-Plane Thermal Transport in Metal-Embedded Graphitic Carbon Nitride

Wenwu Jiang,[1] Hekai Bu,[2] Wei Qiu,[3,*] and Yong Ni[1,*]

[1]*Department of Modern Mechanics, University of Science and Technology of China, Hefei, Anhui 230026, China*

[2]*College of Physical Science and Technology, Bohai University, Jinzhou, China*

[3]*School of Chemistry and The Sackler Center for Computational Molecular and Materials Science, Tel Aviv University, Tel Aviv 6997801, Israel*

[*]Corresponding author.

Email: weiqiu@tauex.tau.ac.il and yni@ustc.edu.cn

## ABSTRACT

Embedding metal atoms within the intrinsic cavities of graphitic carbon nitride (g-$C_3N_4$) modifies its local bonding and vibrational properties, yet its consequences for lattice thermal transport remain poorly understood. Here, we combine a machine-learned potential (MLP) with homogeneous nonequilibrium molecular dynamics (HNEMD) to investigate in-plane thermal transport in Cu-embedded g-$C_3N_4$ across the full range of cavity occupation. The thermal conductivity exhibits a pronounced nonmonotonic dependence on Cu occupation, decreasing at intermediate occupation and partially recovering as the cavities approach full occupation. Frequency-resolved analysis reveals that this behavior originates from the competition between two phonon contributions. The low-frequency vibrations introduced by Cu couple to the intrinsic modes of g-$C_3N_4$, reconstructing the phonon dispersion and suppressing thermal transport below 10 THz. With increasing Cu occupation, the collective mechanical response of the N atoms surrounding the cavities is enhanced, accompanied by increased phonon group velocities in the 20–30 THz range and a corresponding recovery of the high-frequency thermal contribution. The competition between the saturating low-frequency suppression and the increasing high-frequency contribution gives rise to the nonmonotonic thermal conductivity. A similar occupation dependence and spectral redistribution are observed for Ag-embedded g-$C_3N_4$. These results identify competing phonon contributions as the microscopic origin of nonmonotonic thermal transport in metal-embedded g-$C_3N_4$ and suggest that occupation of intrinsic cavities provides a route to redistributing phonon transport in porous crystalline materials.

## 1. Introduction

Graphitic carbon nitride (g-$C_3N_4$) is an intrinsically porous two-dimensional (2D) semiconductor whose periodic nitrogen-rich cavities provide well-defined sites for accommodating isolated metal atoms.[1-3] Coordination of metal atoms with the surrounding nitrogen sites enables the local chemical environment and electronic structure to be modified while largely preserving the underlying C–N framework.[4-8] This structural motif has been widely exploited to tailor charge transfer, adsorption energetics, and catalytic activity, particularly through atomically dispersed transition and coinage metals.[8-12]

While metal occupation has been extensively studied as a means of tuning the electronic and chemical properties of g-$C_3N_4$,[13] its consequences for lattice dynamics remain much less understood. Because the embedded metal atoms are coordinated directly with the pore-edge N atoms, their incorporation alters the local interatomic interactions[14, 15] and can thereby reshape the vibrational spectrum of the host lattice. How these local changes propagate into macroscopic thermal transport, however, has not been established. This issue is also relevant for photoactive g-$C_3N_4$, where photothermal energy conversion can modify the local temperature and can subsequently influence charge-carrier kinetics and thermally activated surface processes.[9, 16]

Previous studies on 2D carbon nitride materials have established that lattice thermal transport is highly sensitive to atomic structure and bonding topology.[17-21] In g-$C_3N_4$, however, most thermal-transport studies have focused on perturbations of the host framework itself, such as strain and defects.[17, 18, 22] Occupation of the intrinsic cavities by foreign atoms represents a distinct form of structural modification: additional atoms and interatomic interactions are introduced within the intrinsic cavities while the topology of the host C–N framework remains largely preserved. Moreover, the fraction of occupied cavities provides a continuous structural variable for systematically examining the influence of embedded species on lattice dynamics. Despite extensive studies of metal-containing g-$C_3N_4$, how lattice thermal transport evolves with cavity occupation remains largely unknown.

In this work, we develop a neuroevolution potential (NEP)[23] and employ homogeneous nonequilibrium molecular dynamics (HNEMD)[24] to systematically investigate in-plane thermal transport in Cu-embedded g-$C_3N_4$ across the full range of cavity occupation. We find that the thermal conductivity varies nonmonotonically with Cu occupation, decreasing at intermediate occupation and partially recovering as the cavities approach full occupation. Frequency-resolved analysis shows that the initial reduction is dominated by suppressed thermal transport below 10 THz, whereas the subsequent recovery arises from an enhanced contribution in the 20–30 THz range. Vibrational analysis attributes these changes to the reconstruction of the low-frequency phonon dispersion induced by Cu vibrations and to enhanced restoring forces within the pore-edge N–N network accompanied by increased phonon group velocities at higher frequencies, respectively. These results identify the competition between low- and high-frequency phonon contributions as the microscopic origin of the nonmonotonic thermal response.

## 2. Model and methods

### *2.1. Model construction and NEP training*

The model system consists of a monolayer g-$C_3N_4$ sheet containing periodically distributed nitrogen-lined intrinsic cavities. The occupation fraction of the embedded metal atoms is defined as:

$$f_{\mathrm{M}} = \frac{N_{\mathrm{M}}}{N_{\mathrm{cav}}}, \tag{1}$$

where $N_{\mathrm{M}}$ is the number of embedded metal atoms and $N_{\mathrm{cav}}$ is the total number of available cavities in the simulation cell. Accordingly, the occupation fraction satisfies $0 \leq f_{\mathrm{M}} \leq 1$, with $f_{\mathrm{M}} = 0$ and $f_{\mathrm{M}} = 1$ corresponding to pristine and fully occupied g-$C_3N_4$, respectively. For each intermediate occupation fraction, the embedded metal atoms were randomly distributed among the available cavities.

To construct the reference dataset for Cu-containing g-$C_3N_4$, four representative occupation states containing one, two, three, and four Cu atoms in the training supercell were considered. For each Cu occupation, ab initio molecular dynamics (AIMD) simulations were performed in the canonical ensemble at 100, 300, 500, 700, and 900 K using the Vienna Ab initio Simulation Package (VASP).[25, 26] Each AIMD trajectory was propagated for 3 ps with a time step of 1 fs using a Nosé-Hoover thermostat, and 25 configurations were extracted at each temperature. This procedure yielded 500 thermally sampled Cu-containing configurations in total. To further sample locally distorted environments beyond those accessed by thermal fluctuations, 30 additional configurations were generated for each Cu occupation by applying random perturbations to the atomic coordinates and lattice parameters, giving another 120 structures. In addition, 200 configurations of pristine g-$C_3N_4$, including both finite-temperature and randomly perturbed structures, were included in the training dataset. The resulting initial dataset therefore contains 820 configurations, comprising 620 Cu-containing and 200 pristine g-$C_3N_4$ structures. The overall workflow for the construction of the training dataset is summarized in **Fig. 1**.

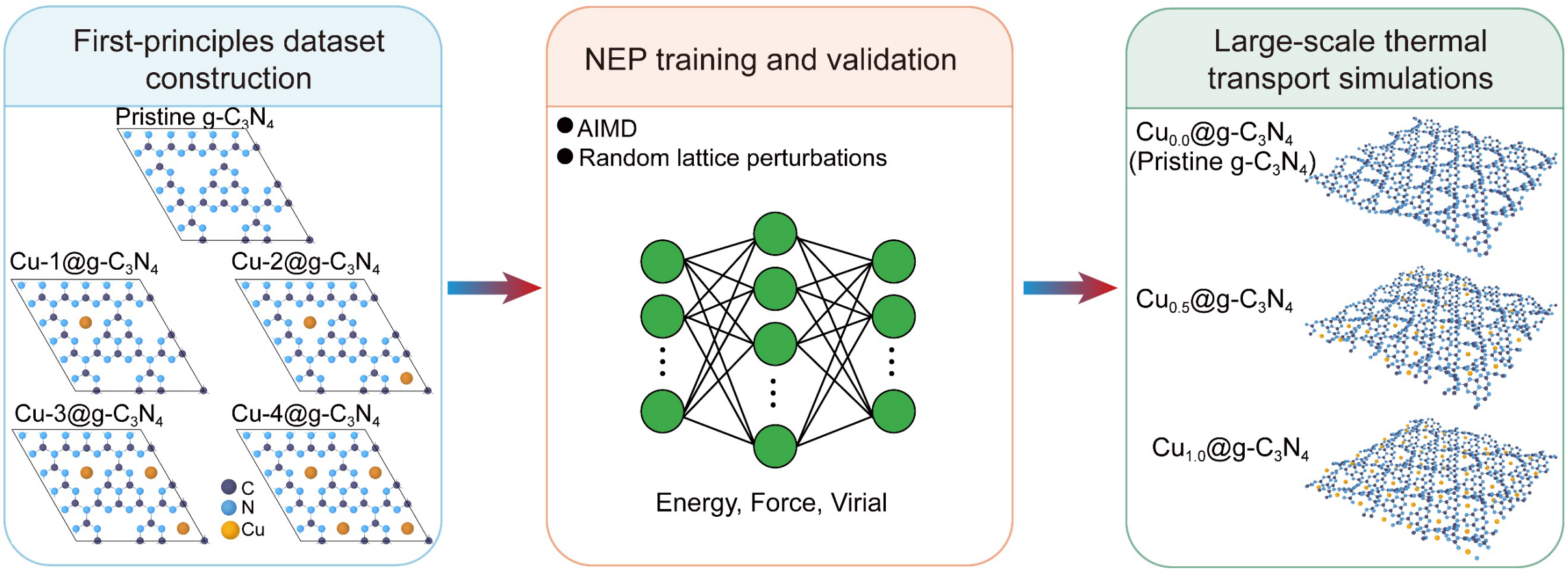

**Fig. 1.** Multiscale computational framework for Cu-embedded g-$C_3N_4$. First-principles reference structures were constructed from pristine g-$C_3N_4$ and representative Cu occupation states (Cu-1@ g-$C_3N_4$ to Cu-4@ g-$C_3N_4$), where the numbers 1–4 denote the number of Cu atoms in the training supercell. Gray, blue, and yellow spheres represent C, N, and Cu atoms, respectively. The configurational space was further sampled by AIMD at 100–900 K and by random atomic and lattice perturbations. These reference data were used to train and validate the NEP against first-principles energies, forces, and virials. The trained NEP was then employed in large-scale thermal-transport simulations for pristine ($f_{\mathrm{Cu}} = 0$), partially occupied $Cu_{0.5}$@ g-$C_3N_4$ ($f_{\mathrm{Cu}} = 0.5$), and fully occupied $Cu_{1.0}$@ g-$C_3N_4$ ($f_{\mathrm{Cu}} = 1$) systems.

First-principles calculations were subsequently performed to obtain the reference energies, atomic forces, and virial tensors used for NEP training. All calculations were carried out using

VASP within the projector augmented-wave (PAW)[27] formalism. Exchange and correlation were treated using the Perdew-Burke-Ernzerhof (PBE)[28] functional within the generalized-gradient approximation. A plane-wave kinetic-energy cutoff of 850 eV was employed. The Brillouin zone was sampled with a Γ-centered $k$-point grid corresponding to a reciprocal-space spacing of 0.15 $Å^{-1}$, and the electronic self-consistent convergence criterion was set to $10^{-7}$ eV. A vacuum spacing of 20 Å was introduced along the out-of-plane direction to minimize spurious interactions between periodically repeated monolayers.

The interatomic interactions in Cu-embedded g-$C_3N_4$ were described using the fourth-generation neuroevolution potential (NEP4)[29] framework. Within this formalism, the site energy $U_i$ of atom $i$ is expressed by a feedforward neural network with a single hidden layer containing $N_{\mathrm{neu}}$ neurons:[23]

$$U_i = \sum_{\mu=1}^{N_{\mathrm{neu}}} w_{\mu}^{(1)} \tanh\left(\sum_{\nu=1}^{N_{\mathrm{des}}} w_{\mu\nu}^{(0)} q_{\nu}^{i} + b_{\mu}^{(0)}\right) + b^{(1)}, \quad (2)$$

where $q_{\nu}^{i}$ denotes the $N_{\mathrm{des}}$-dimensional descriptor vector that characterizes the local atomic environment of atom $i$, and $\tanh(x)$ is the activation function. The quantities $w^{(0)}$, $w^{(1)}$, $b^{(0)}$, and $b^{(1)}$ are the trainable weights and biases of the neural network. The local descriptor set includes both radial and angular contributions. The model parameters were optimized by minimizing a loss function comprising the weighted root-mean-square errors (RMSEs) of energies, forces, and virials, along with regularization terms. Details of the hyperparameters used for NEP training are provided in Sec. 1 of the Supporting Information (SI). The cutoff radii for both the radial and angular descriptors were set to 5 Å, beyond which the energy, force, and virial errors showed only marginal improvement, as demonstrated by the convergence tests presented in Sec. 2 of the SI. For each descriptor type, nine radial basis functions were employed, with each basis function further expanded as a linear combination of 13 Chebyshev polynomials. Following the atomic cluster expansion strategy for the construction of angular descriptors, three-body and four-body correlations described by spherical harmonics were included up to angular orders $l = 4$ and $l = 2$, respectively. The final NEP model was trained for 500,000 generations using a full-batch optimization scheme.

*2.2. Validation of the NEP*

The accuracy of the trained NEP was first assessed by comparing the predicted energies, atomic forces, and virial tensors with the corresponding first-principles reference values. As shown in **Fig. 2**, the NEP predictions closely follow the first-principles data for both the training and test sets over the sampled configurational space. The energy RMSEs are 1.41 and 1.39 meV/atom for the training and test sets, respectively, while the corresponding force RMSEs are 146.10 and 149.67 meV/Å. For the virial, RMSEs of 13.10 and 13.56 meV/atom are obtained for the training and test sets, respectively. The comparable errors between the two datasets indicate that the trained potential retains similar accuracy for configurations not included in the fitting procedure.

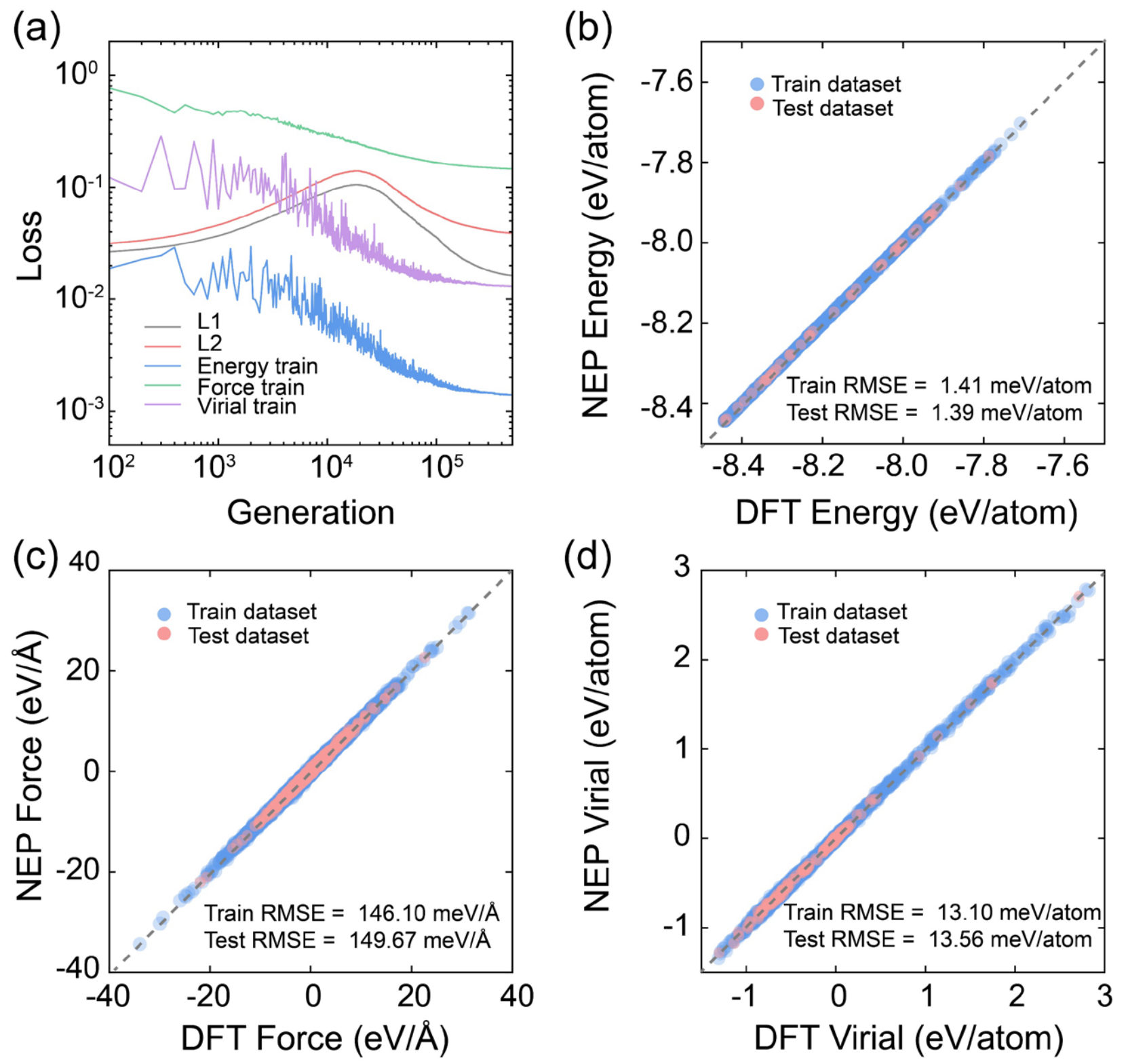

**Fig. 2.** (a) Evolution of the energy, force, and virial loss components during the training process as a function of training generations (steps). Comparison of (b) energy, (c) force, and (d) virial calculated by NEP against DFT reference data for the training and test datasets.

Because the present work focuses on phonon-mediated thermal transport, agreement in energies and forces alone is insufficient to establish the suitability of the potential. We therefore further examined quantities directly related to lattice dynamics. The phonon dispersion relations predicted by the NEP model are compared with those obtained from PBE calculations (see Sec. 3 of the SI). The NEP reproduces the first-principles phonon branches over the investigated frequency range. In addition, the radial distribution functions calculated from NEP and first-principles molecular dynamics exhibit close agreement at both 300 and 500 K, as presented in Sec. 4 of the SI, demonstrating that the potential captures the finite-temperature structural environments sampled by the system.

*2.3. HNEMD method*

We use the efficient HNEMD method to calculate the thermal conductivity of Cu-embedded g-$C_3N_4$. In this method, the system is driven out of equilibrium by an external field, and the corresponding external driving force acting on atom $i$ is written as:[24]

$$\boldsymbol{F}_i^{\mathrm{ext}} = E_i \boldsymbol{F}_e + \boldsymbol{F}_e \cdot \mathbf{W}_i, \tag{3}$$

where $\boldsymbol{F}_e$ is a vectorial driving force parameter with units of inverse length, $E_i$ is the total energy of atom $i$, and $\mathbf{W}_i$ is the virial tensor of atom $i$, defined as:

$$\mathbf{W}_i = \sum_{j \neq i} \left( \frac{\partial U_j}{\partial \boldsymbol{r}_{ji}} \otimes \boldsymbol{r}_{ij} \right). \tag{4}$$

Here, $U_i$ is the potential energy assigned to atom $i$, and $\boldsymbol{r}_{ij} \equiv \boldsymbol{r}_j - \boldsymbol{r}_i$, with $\mathbf{r}_i$ being the position vector of atom $i$. Under the action of the external field, the system reaches a nonequilibrium steady-state, in which the ensemble-averaged heat current $\langle \boldsymbol{J} \rangle$ satisfies

$$\frac{\langle J^\alpha \rangle}{TV} = \sum_\beta \kappa^{\alpha\beta} F_e^\beta, \tag{5}$$

where $T$ and $V$ denote the system temperature and volume, respectively, and $\kappa^{\alpha\beta}$ is the $\alpha\beta$ component of the thermal conductivity tensor. The Cu-embedded g-$C_3N_4$ system exhibits only weak in-plane anisotropy, with comparable thermal conductivities along the $x$ (zigzag) and $y$ (armchair) directions (see Sec. 5 of the SI). Accordingly, the $x$-direction component is adopted as representative of the in-plane thermal transport, and $\kappa^{xx}$ is evaluated as:

$$\kappa^{xx} = \frac{\langle J^x \rangle}{TVF_{\mathrm{e}}^x}. \tag{6}$$

The driving-field strength $F_e^x$ was set to $5 \times 10^{-4}/\text{Å}$ for Cu-embedded g-$C_3N_4$. An important advantage of the HNEMD method is that it allows a spectral decomposition of the thermal conductivity. Specifically, the in-plane thermal conductivity along the $x$ direction can be expressed as:

$$\kappa^{xx} = \int_0^{\infty} \frac{\mathrm{d}\omega}{2\pi} \kappa^{xx}(\omega), \tag{7}$$

where the frequency-resolved thermal conductivity is given by

$$\kappa^{xx}(\omega) = \frac{2}{TVF_e^x} \int_{-\infty}^{\infty} \mathrm{d}t\, \mathrm{e}^{i\omega t} K^x(t). \tag{8}$$

Here, $K^x(t)$ is the $x$-component of the virial–velocity correlation function vector $\boldsymbol{K}(t)$, defined as:

$$\boldsymbol{K}(t) = \sum_i \langle \mathbf{W}_i(0) \cdot \boldsymbol{v}_i(t) \rangle, \tag{9}$$

where $\boldsymbol{v}_i(t)$ is the velocity of atom $i$ at time $t$.

All HNEMD calculations were performed using the GPUMD package.[29] The initial simulation cells of Cu-embedded g-$C_3N_4$ were constructed with in-plane dimensions of approximately 30.0 × 29.7 $nm^2$. A vacuum spacing of 100 Å was introduced along the out-of-plane ($z$) direction to minimize spurious interactions between periodic images. The convergence of the calculated thermal conductivity with respect to the lateral cell size was independently verified, as detailed in Sec. 6 of the SI. Each system was first equilibrated for 200 ps in the isothermal-isobaric (NPT)[30] ensemble at 300 K and zero in-plane pressure, followed by an additional 200 ps equilibration in the canonical (NVT) ensemble. A subsequent 5 ns HNEMD production run was then performed in the NVT ensemble with the external driving force applied throughout the production stage. The NPT equilibration was carried out using the Berendsen method, whereas the NVT ensemble was maintained with a Nosé-Hoover chain thermostat. For the conversion of the two-dimensional heat current to thermal conductivity, an effective layer thickness of 3.3 Å was adopted.[3] The influence of long-range dispersion interactions was additionally examined by including a D3 correction.[31] The resulting thermal conductivities for representative Cu occupation fractions differ only marginally from those obtained without the dispersion correction, indicating that long-range van der Waals interactions have a negligible influence on the in-plane thermal transport considered here (see

Sec. 7 of the SI). The thermal conductivity was confirmed to be well converged within the production time under steady-state conditions (see Sec. 6 of the SI). Three independent trajectories were performed for each condition, and the reported thermal conductivity and its statistical uncertainty were obtained from the corresponding ensemble average.

*2.4. Phonon SED and DOS calculations*

To characterize the changes in phonon dispersion induced by Cu embedding in monolayer g-$C_3N_4$, we computed the phonon spectral energy density (SED)[32, 33] as a function of wave vector $\boldsymbol{k}$ and angular frequency $\omega$. In molecular dynamics simulations, the SED is evaluated as:[33, 34]

$$\Phi(\boldsymbol{k},\omega) = \frac{1}{4\pi\tau_0 N}\sum_{\alpha}^{\{x,y,z\}}\sum_{b}^{B} m_b \times \left|\int_0^{\tau_0}\sum_{n_{x,y,z}}^{N} v_\alpha\left(\begin{smallmatrix} n_{x,y,z} \\ b \end{smallmatrix}; t\right) \times \exp\left[i\boldsymbol{k}\cdot\boldsymbol{r}\left(\begin{smallmatrix} n_{x,y,z} \\ 0 \end{smallmatrix}\right) - i\omega t\right]\mathrm{d}t\right|^2. \quad (10)$$

Here, $\tau_0$ denotes the total simulation time, $N$ is the number of unit cells included in the sampled supercell, and $b$ labels the atoms within an individual unit cell. The quantity $m_b$ represents the mass of atom $b$, while $n_{x,y,z}$ identifies the unit cell along the $x$, $y$, and $z$ directions. The term $v_\alpha\left(\begin{smallmatrix} n_{x,y,z} \\ b \end{smallmatrix}; t\right)$ is the $\alpha$-component of the velocity of atom $b$ in unit cell $n$ at time $t$, and $\boldsymbol{r}\left(\begin{smallmatrix} n_{x,y,z} \\ 0 \end{smallmatrix}\right)$ denotes its equilibrium position.

The molecular dynamics simulations used for the SED calculations were performed using LAMMPS package.[35] Cu-embedded g-$C_3N_4$ structures were replicated into $25 \times 25 \times 1$ supercells to provide sufficient wave-vector resolution. Each system was first equilibrated for 100 ps in the NPT ensemble at 300 K with zero in-plane pressure and a fixed out-of-plane cell dimension, followed by an additional 100 ps equilibration in the NVT ensemble. A subsequent 250 ps production run was performed in the NVE ensemble using a time step of 0.5 fs. During the NVE trajectory, atomic positions and velocities were sampled every 2.5 fs for subsequent SED analysis. The frequency-resolved SED spectra were then calculated using the PYSED package.[33]

The phonon density of states (DOS) was calculated from the Fourier transform of the velocity autocorrelation function (VACF). For each atomic species $s$, the phonon DOS was evaluated as:[36]

$$g_s(\omega) = \frac{1}{N_s} \sum_{i \in s} \left( \int_{-\infty}^{+\infty} \langle \mathbf{v}_i(0) \cdot \mathbf{v}_i(\mathrm{t}) \rangle \, e^{i\omega t} \mathrm{d}t \right) \tag{11}$$

where $N_s$ is the number of atoms of species $s$, $\mathbf{v}_i(t)$ is the velocity vector of atom $i$ at time $t$, and $\langle \cdot \rangle$ denotes the ensemble average.

For the phonon DOS calculations, each system was first initialized at 300 K and equilibrated for 500 ps in the NPT ensemble at 300 K and zero pressure, followed by an additional 500 ps equilibration in the NVT ensemble using GPUMD package.[29] A time step of 1 fs was used throughout the simulations. The production trajectory was then generated from a 500 ps NVE simulation, during which the VACFs were accumulated for the subsequent phonon DOS analysis.

## 3. Results and discussion

### *3.1. Nonmonotonic thermal transport and spectral redistribution with Cu occupation*

**Figure 3**a shows the in-plane thermal conductivity of Cu-embedded g-$C_3N_4$ as a function of the cavity occupation fraction $f_{\mathrm{Cu}}$. At 300 K, pristine g-$C_3N_4$ exhibits a thermal conductivity of 6.19 ± 0.04 $\mathrm{Wm^{-1}K^{-1}}$. The calculated value is also consistent with the broad range of in-plane thermal conductivities reported for monolayer g-$C_3N_4$ in previous MD studies, which span approximately 1.5–14.1 $\mathrm{Wm^{-1}K^{-1}}$ depending on the interatomic potential and simulation protocol.[22, 37-39] This agreement provides an additional check on the physical reasonableness of the present thermal-transport calculations. As Cu atoms are progressively introduced into the intrinsic cavities, the thermal conductivity decreases substantially and reaches a broad minimum at intermediate occupation, approximately within $f_{\mathrm{Cu}} = 0.5 \sim 0.7$, where $\kappa$ is about 4.15–4.18 $\mathrm{Wm^{-1}K^{-1}}$. This corresponds to a reduction of approximately 33% relative to the pristine system. Upon further increasing the Cu occupation, the thermal conductivity reverses its trend and gradually recovers, reaching 4.60 ± 0.01 $\mathrm{Wm^{-1}K^{-1}}$ at full occupation ($f_{\mathrm{Cu}} = 1$). Thus, Cu incorporation does not produce a monotonic suppression of thermal transport; instead, the thermal conductivity exhibits a pronounced minimum at intermediate cavity occupation. Moreover, a similar decrease-and-recovery behavior is observed over the investigated temperature range of 200–500 K. Although the magnitude of the thermal conductivity

decreases with increasing temperature, the characteristic occupation dependence is retained, with the minimum consistently occurring at intermediate Cu occupation.

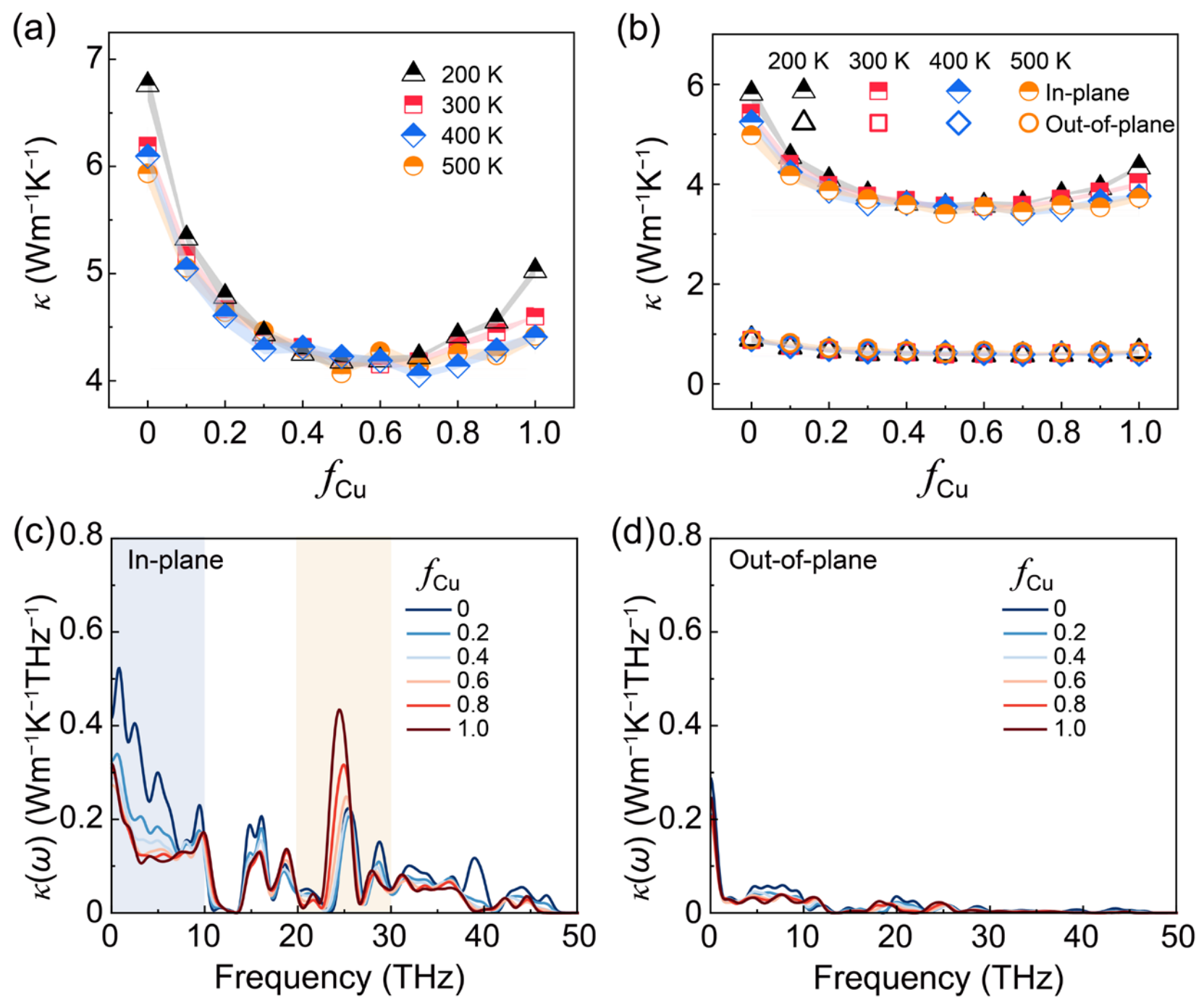

**Fig. 3.** Cu-occupation dependence of phonon thermal transport in monolayer Cu-embedded g-$C_3N_4$. (a) In-plane thermal conductivity ($\kappa$) as a function of the Cu occupation fraction ($f_{\mathrm{Cu}}$) at 200–500 K. (b) In-plane (filled symbols) and out-of-plane (open symbols) contributions to the thermal conductivity at 200–500 K. (c) Frequency-resolved in-plane thermal conductivity $\kappa(\omega)$ for representative Cu occupation fractions at 300 K. The shaded regions highlight the low-frequency (0–10 THz) and high-frequency (20–30 THz) ranges associated with the dominant occupation-dependent changes in thermal transport. (d) Frequency-resolved out-of-plane thermal conductivity $\kappa(\omega)$ for representative Cu occupation fractions at 300 K.

To determine which component of lattice motion is primarily responsible for this nonmonotonic response, we further decompose the thermal conductivity into in-plane and out-of-plane contributions. As shown in **Fig. 3**b, the in-plane component dominates the thermal transport over the entire range of Cu occupation, whereas the out-of-plane contribution remains substantially smaller. More importantly, the in-plane contribution retains the pronounced decrease-and-recovery dependence on $f_{\mathrm{Cu}}$, while the out-of-plane component varies only weakly with cavity occupation. The nonmonotonic variation of the total in-plane thermal conductivity therefore originates predominantly from changes in thermal transport associated with in-plane atomic motion.

Further insight is obtained from the frequency-resolved thermal conductivity. As shown in **Fig. 3**c, the in-plane spectral thermal conductivity below 10 THz decreases markedly after Cu incorporation, whereas the contribution in the 20–30 THz range increases with increasing Cu occupation. The corresponding out-of-plane spectral thermal conductivity is shown in **Fig. 3**d. Its magnitude is considerably smaller than that of the in-plane contribution over nearly the entire frequency range and changes only modestly with Cu occupation. In particular, no high-frequency enhancement comparable to that observed in the in-plane spectrum is present. These results further confirm that the nonmonotonic thermal response is governed primarily by the redistribution of in-plane thermal transport.

To quantify the evolution of the two frequency ranges with increasing Cu occupation, we evaluate the change in their thermal-conductivity contributions between successive occupation fractions. Since $f_{\mathrm{Cu}}$ increases in steps of $\Delta f = 0.1$, the incremental loss of the low-frequency (0–10 THz) contribution is defined as:

$$\Delta\kappa_{0-10\,\mathrm{THz}}(f_{\mathrm{Cu}}) = \kappa_{0-10\,\mathrm{THz}}(f_{\mathrm{Cu}} - 0.1) - \kappa_{0-10\,\mathrm{THz}}(f_{\mathrm{Cu}}), \quad (12)$$

whereas the change in the high-frequency (20–30 THz) contribution is defined as:

$$\Delta\kappa_{20-30\,\mathrm{THz}}(f_{\mathrm{Cu}}) = \kappa_{20-30\,\mathrm{THz}}(f_{\mathrm{Cu}}) - \kappa_{20-30\,\mathrm{THz}}(f_{\mathrm{Cu}} - 0.1), \quad (13)$$

Using Eqs. (12) and (13), the incremental changes in the two spectral contributions are summarized in **Fig. 4**. The initial introduction of Cu produces a large loss of the 0–10 THz contribution, whereas the additional low-frequency suppression decreases rapidly with increasing $f_{\mathrm{Cu}}$ and becomes small at high occupation. In contrast, the 20–30 THz contribution initially changes only weakly but becomes increasingly positive as the cavities are filled. At $f_{\mathrm{Cu}} = 0.5 \sim 0.7$, the additional low-frequency loss becomes comparable to the high-frequency gain, so that their competing contributions nearly balance each other. This crossover gives rise to the broad minimum in the total thermal conductivity observed around $f_{\mathrm{Cu}} = 0.5 \sim 0.7$. With further Cu occupation, the low-frequency suppression approaches saturation, whereas the high-frequency contribution continues to increase and eventually exceeds the remaining low-frequency loss, leading to the recovery of the total thermal conductivity. The nonmonotonic dependence of $\kappa$ on Cu occupation therefore results from the changing balance between these two spectral contributions.

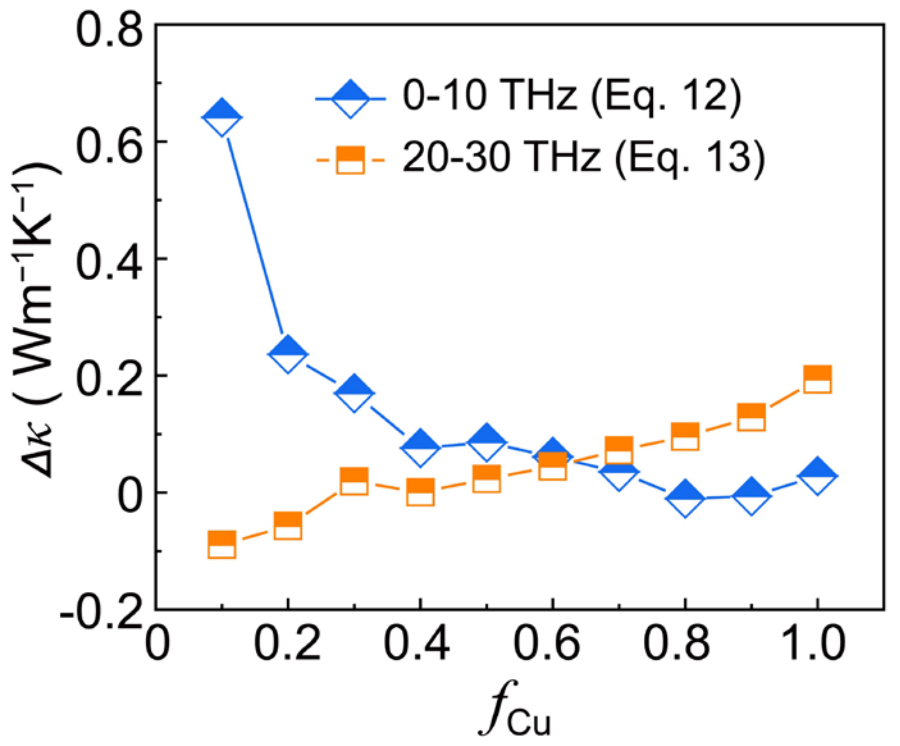

**Fig. 4.** Incremental changes in the spectral thermal-conductivity contributions between successive Cu occupation fractions at 300 K, calculated using Eqs. (12) and (13). Blue diamonds denote the loss of the low-frequency contribution in the 0–10 THz range, whereas orange squares denote the change in the high-frequency contribution in the 20–30 THz range.

### *3.2. Suppression of the low-frequency phonon contribution by dispersion reconstruction*

To elucidate the origin of the suppressed thermal-conductivity contribution below 10 THz, we first examine the vibrational characteristics introduced by Cu embedding. **Fig. 5**a compares the in-plane PDOS of the g-$C_3N_4$ host with the Cu-projected contribution. The Cu vibrational states are concentrated predominantly below approximately 10 THz and overlap with the intrinsic low-frequency modes of g-$C_3N_4$. This frequency coincidence indicates that the embedded Cu atoms can directly perturb the same low-frequency phonon modes that dominate the reduction in spectral thermal conductivity shown in **Fig. 3**c.

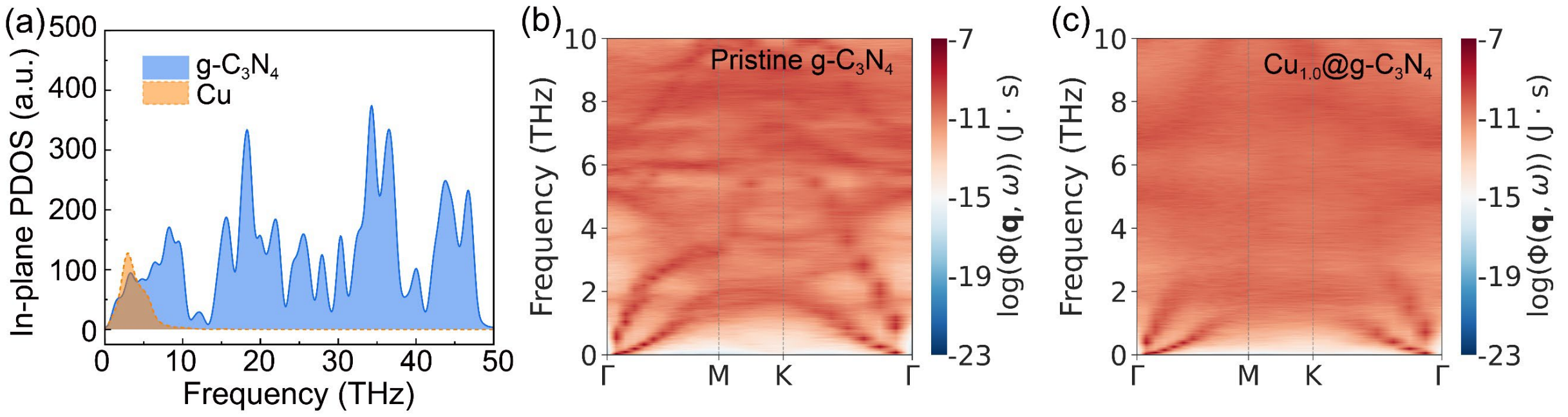

**Fig. 5.** Cu-induced modification of low-frequency phonon dynamics in g-$C_3N_4$. (a) In-plane PDOS of pristine g-$C_3N_4$ together with the Cu-projected contribution. (b,c) Phonon SED along the Γ-M-K-Γ path for pristine g-$C_3N_4$ and fully occupied $Cu_{1.0}$@g-$C_3N_4$, respectively.

To further determine how Cu embedding modifies these heat-carrying modes, we analyze the phonon SED. **Figs. 5**b,c show the low-frequency SED of pristine g-$C_3N_4$ and fully occupied $Cu_{1.0}$@g-$C_3N_4$, respectively. Compared with the pristine system, Cu embedding produces pronounced broadening and a loss of definition of the low-frequency phonon branches, features that are generally associated with shortened phonon lifetimes and enhanced phonon scattering,[40-42] indicating a substantial perturbation of the intrinsic phonon dispersion. The correspondence between this spectral broadening and the reduced thermal-conductivity contribution below 10 THz suggests that the low-frequency suppression originates from Cu-induced disruption of the intrinsic heat-carrying phonons.

### *3.3. Recovery of the high-frequency phonon contribution through enhanced pore-edge coupling*

In contrast to the low-frequency suppression discussed above, the thermal-conductivity contribution in the 20–30 THz range increases progressively at high Cu occupation (see **Fig. 3**c). To identify the origin of this recovery, we examine the lattice dynamics in the corresponding frequency range. **Figs. 6**a,b compare the phonon SED of pristine g-$C_3N_4$ and fully occupied $Cu_{1.0}$@g-$C_3N_4$, respectively. Cu occupation produces a pronounced reorganization of the phonon branches between 20 and 30 THz, particularly around 24–26 THz, where the enhancement of the spectral thermal conductivity is most evident. Unlike the broadening observed for the low-frequency modes, the high-frequency branches remain well defined, indicating that the recovery of thermal transport in this frequency range is associated with a modification of their propagation characteristics.

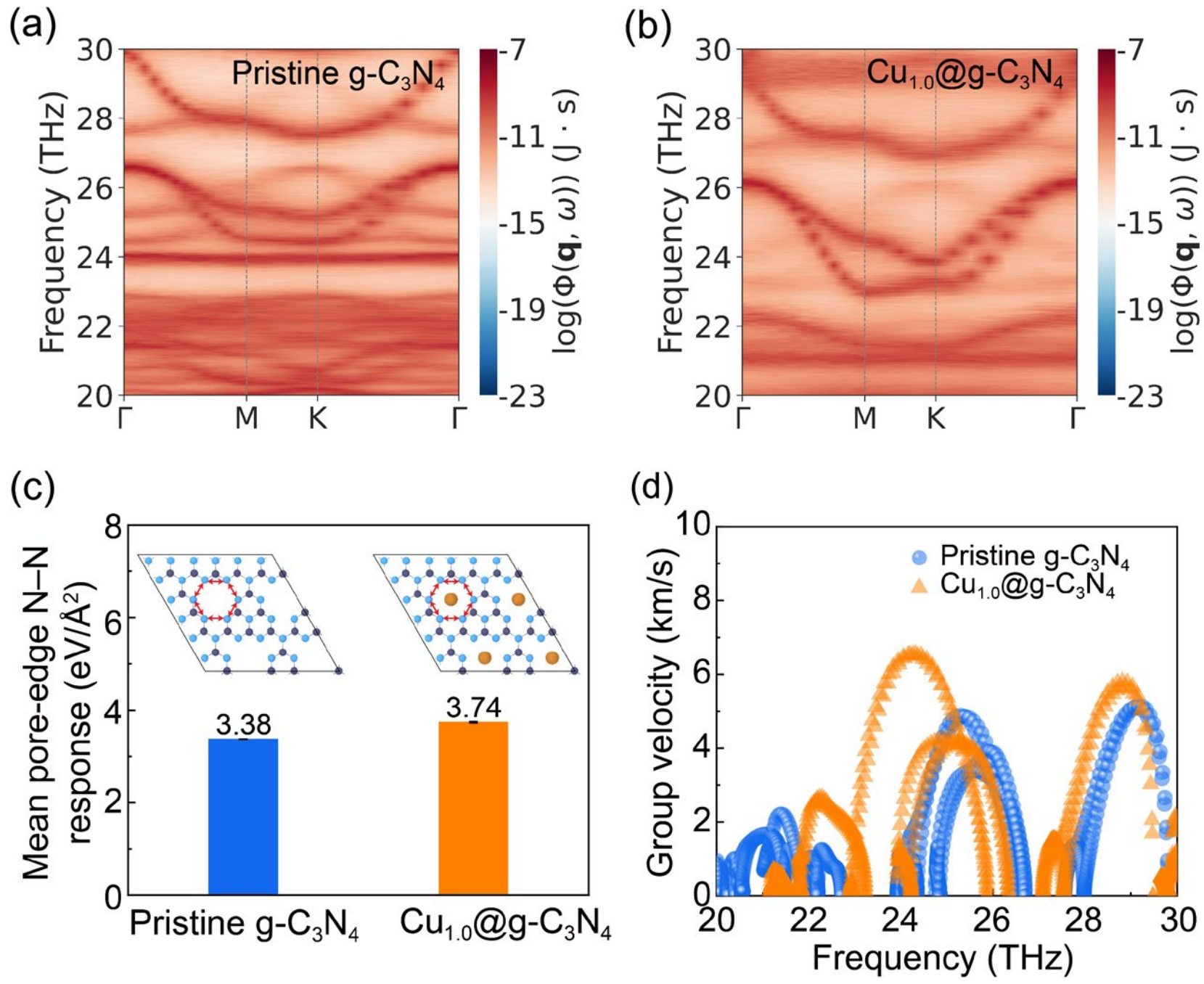

**Fig. 6.** Cu-induced modification of high-frequency phonon transport in g-$C_3N_4$. (a,b) Phonon SED in the 20–30 THz range along the Γ-M-K-Γ path for pristine g-$C_3N_4$ and $Cu_{1.0}$@g-$C_3N_4$, respectively. (c) Mean force response of the pore-edge N–N network, increasing from 3.38 eV/Å$^2$ in pristine g-$C_3N_4$ to 3.74 eV/Å$^2$ after full Cu occupation. (d) Phonon group velocities in the 20–30 THz range for the pristine and fully occupied systems.

To isolate the direct mechanical effect of Cu on the N atoms surrounding the cavity, we analyzed the local force response using finite displacements. The relaxed fully occupied structure was taken as the reference. A second structure was generated by removing all Cu atoms while keeping the simulation cell and all C/N coordinates unchanged. For each structure, the six N atoms surrounding the selected cavity were considered in turn, as shown in **Fig. 6**c. Each N atom was displaced individually by ±0.005 Å along the $x$ and $y$ directions, while all other atoms were kept at their reference positions. The resulting atomic forces were used to obtain the in-plane force constants through central differences: $\Phi_{j\alpha,i\beta} =$ $-\frac{F_{j\alpha}(+\Delta u_{i\beta})-F_{j\alpha}(-\Delta u_{i\beta})}{2\Delta u}$, where $\alpha, \beta \in \{x, y\}$. For each atomic pair $i$ and $j$, the four components were assembled into a $2 \times 2$ in-plane force-constant block, and its Frobenius norm, $C_{ji}^{\text{in-plane}} = \|\Phi_{ji}^{xy}\|_{\text{F}}$, was used to characterize the magnitude of the in-plane force response. For a displaced pore-edge N atom $i$, the cooperative response of the other five N atoms surrounding the same cavity was defined as $\sum_{j \neq i} C_{ji}^{\text{in-plane}}$, and the reported force response of the pore-edge N–N network was obtained by averaging over the six N sites.

As shown in **Fig. 6**c, this pore-edge N response increases from 3.38 to 3.74 eV/Å$^2$ when Cu is retained in the fixed geometry, demonstrating that Cu–N interactions enhance the cooperative mechanical response among the N atoms surrounding the same cavity. This enhancement indicates a redistribution of the local response toward the pore-edge N network. For further comparison, the phonon group velocities were calculated using Phonopy package[43] from second-order force constants obtained with the trained NEP. As shown in **Fig. 6**d, the group velocities in the 20–30 THz range increase after Cu embedding. The combined results therefore support a picture in which Cu–N coordination reorganizes the collective dynamics of the pore-edge N atoms, modifies the C/N-dominated high-frequency phonon modes, and increases their group velocities. The resulting enhancement of the high-frequency contribution partially compensates for the suppressed low-frequency transport and accounts for the recovery of the total thermal conductivity at high Cu occupation.

### *3.4. Nonmonotonic thermal transport with Ag occupation*

To examine whether the nonmonotonic thermal response is specific to Cu, we further consider Ag occupation of the intrinsic cavities in g-$C_3N_4$. The NEP for Ag-embedded g-$C_3N_4$ was trained using the same computational settings and training protocol as those adopted for the Cu system, and the corresponding training results are provided in Sec. 8 of the SI. As shown in **Fig. 7**a, the in-plane thermal conductivity at 300 K exhibits a similar nonmonotonic dependence on the Ag occupation fraction $f_{\text{Ag}}$. The thermal conductivity decreases as Ag atoms are introduced, reaches a broad minimum at intermediate occupation, and then partially recovers as the cavities approach full occupation. The decomposition in **Fig. 7**b further shows that the in-plane contribution dominates over the entire occupation range and reproduces the same decrease-and-recovery behavior, whereas the out-of-plane contribution remains much smaller and varies only weakly with $f_{\mathrm{Ag}}$.

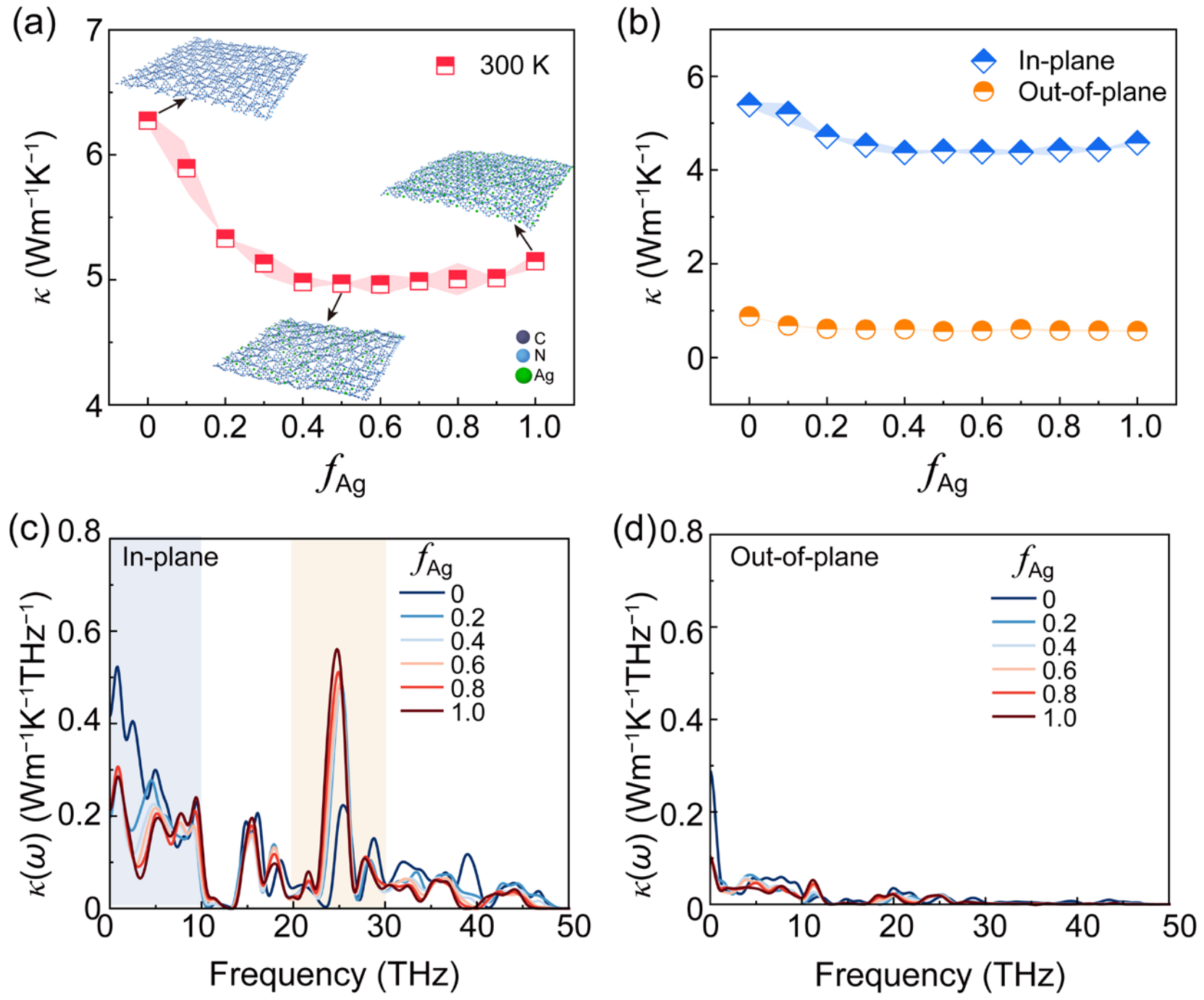

**Fig. 7.** Thermal transport in Ag-embedded g-$C_3N_4$ at 300 K. (a) In-plane thermal conductivity ($\kappa$) as a function of the Ag occupation fraction $f_{\mathrm{Ag}}$. Insets show representative atomic structures at selected occupation fractions, where gray, blue, and green spheres denote C, N, and Ag atoms, respectively. (b) In-plane and out-of-plane contributions to the thermal conductivity as functions of $f_{\mathrm{Ag}}$. (c) Frequency-resolved in-plane thermal conductivity $\kappa(\omega)$ for representative Ag occupation fractions. The shaded regions indicate the low-frequency (0–10 THz) and high-frequency (20–30 THz) ranges. (d) Corresponding frequency-resolved out-of-plane thermal conductivity.

The frequency-resolved results reveal a similar redistribution of the heat-carrying phonon contributions. As shown in **Fig. 7**c, Ag occupation suppresses the in-plane spectral thermal conductivity below 10 THz, while the contribution in the 20–30 THz range increases with increasing $f_{\mathrm{Ag}}$. The corresponding out-of-plane spectral contribution remains comparatively small over the entire frequency range (see **Fig. 7**d). These results show that the nonmonotonic thermal conductivity of Ag-embedded g-$C_3N_4$ is accompanied by the same competition between suppressed low-frequency and enhanced high-frequency contributions identified for Cu, which indicates that this spectral competition may represent a more general feature of metal occupation in porous g-$C_3N_4$.

**4. Conclusion**

In summary, we investigated the effect of metal occupation on in-plane lattice thermal transport in porous g-$C_3N_4$ using a neuroevolution potential combined with HNEMD simulations. Cu occupation produces a nonmonotonic thermal response, with the thermal conductivity decreasing at intermediate occupation and partially recovering as the cavities approach full occupation. Frequency-resolved analysis shows that this behavior arises from competing phonon contributions: Cu embedding suppresses thermal transport below 10 THz by strongly perturbing the intrinsic low-frequency modes, while increasing Cu occupation enhances the 20–30 THz contribution through a reorganization of the pore-edge N dynamics accompanied by larger phonon group velocities. The competition between these two spectral responses gives rise to the nonmonotonic dependence of the total thermal conductivity. Ag-embedded g-$C_3N_4$ exhibits a similar occupation dependence and spectral redistribution, indicating that this behavior is not specific to Cu. These results demonstrate that occupation of intrinsic cavities can redistribute thermal transport across different parts of the phonon spectrum and thereby provide a general microscopic route for tuning lattice thermal transport in porous crystalline materials.

## Supporting Information

Supporting Information contains the following sections: Hyperparameters for NEP training, convergence of NEP prediction errors with cutoff radius, comparison of phonon dispersions calculated by PBE and NEP, radial distribution functions at 300 and 500 K, in-plane thermal conductivity along zigzag and armchair directions, convergence of HNEMD thermal conductivity, effect of the D3 dispersion correction, and training and validation of the NEP for Ag-embedded g-$C_3N_4$.

**Author Contributions**

Y.N. and W.Q. conceived the research and supervised the project. W.J., W.Q. and H.B. performed all MD simulations. All authors contributed to data analysis and manuscript writing.

**Notes**

The authors declare no competing financial interest.

**Acknowledgments**

Y.N. acknowledges the National Key Research and Development Program of China (Grant No. 2022YFA1203602) and the National Natural Science Foundation of China (Grant No. 42488301). W.J. acknowledges the Fundamental Research Funds for the Central Universities. W.Q. acknowledges the postdoctoral fellowships of the Sackler Center for Computational Molecular and Materials Science and the Ratner Center for Single Molecule Science at Tel Aviv University. Computations were conducted at the National Supercomputer TianHe-1(A) Center in Tianjin.

**Data availability**

All input files associated with the training and validation of the NEP are freely available through Zenodo at: https://doi.org/10.5281/zenodo.22171016.